# Chiral-Index Resolved Length Mapping of Carbon Nanotubes in Solution Using Electric-Field Induced Differential Absorption Spectroscopy

Wenshan Li[1,2], Frank Hennrich[1], Benjamin S. Flavel[1], Manfred M. Kappes[1,3], and Ralph Krupke[1,2*]

[1]Institute of Nanotechnology, Karlsruhe Institute of Technology, 76021 Karlsruhe, Germany
[2]Department of Materials and Earth Sciences, Technische Universität Darmstadt, 64287 Darmstadt, Germany
[3]Institute of Physical Chemistry, Karlsruhe Institute of Technology, 76128 Karlsruhe, Germany

email: wenshan.li@kit.edu; krupke@kit.edu

## ABSTRACT

The length of single-walled carbon nanotubes (SWCNTs) is an important metric for the integration of SWCNTs into devices and for the performance of SWCNT-based electronic or optoelectronic applications. In this work we propose a rather simple method based on electric-field induced differential absorption spectroscopy (EFIDAS) to measure the chiral-index-resolved average length of SWCNTs in dispersions. The method takes advantage of the electric-field induced length-dependent dipole moment of nanotubes and has been verified and calibrated by atomic force microscopy. This method not only provides a low cost, *in-situ* approach for length measurements of SWCNTs in dispersion, but due to the sensitivity of the method to the SWCNT chiral index, the chiral index dependent average length of fractions obtained by chromatographic sorting can also be derived. Also, the determination of the chiral-index resolved length distribution seems to be possible using this method.

Single-walled carbon nanotubes (SWCNTs) have a unique structure-property correlation which allows the nanotube's chiral index and hence its diameter to be determined by spectroscopic methods such as absorption, fluorescence or Raman spectroscopy[1]. In contrast, the length of a carbon nanotube has little influence on its optical properties and therefore a direct spectroscopic access to the SWCNT length beyond 100 nm is not available,[2,3] Consequently, methods that were proven to be very accurate in gauging nano-scale and micro-scale dimensions such as Transmission Electron Microscopy, Scanning Electron Microscopy and Atomic Force Microscopy (AFM) have been applied to measure the length of a nanotube or the length distribution of a nanotube ensemble. In particular the AFM method has evolved as a kind of standard approach to measure nanotube length distributions,[4,5] even though the characterization method requires specific substrates, and additional preparation steps and is associated with uncertainties due to nanotube bundling on surfaces or selective adsorption. Therefore, the search for new SWCNT length measurement methods is ongoing, and especially the field of SWCNT liquid phase sorting would benefit from a fast and practical *in-situ* length characterization method. In recent years, a range of *in-situ* methods have been explored such as dynamic light scattering,[6–9] electrospray differential mobility analysis,[10] diffusional trajectory,[11] shear-aligned photoluminescence anisotropy[12] and analytical ultracentrifugation[13]. The most advanced *in-situ* measurement of the length distribution of dispersed SWNCTs has been reported by the Weisman group who introduced the length analysis by nanotube diffusion method (LAND).[11] The LAND method yields length distributions very similar to the comparative *ex-situ* AFM method, but unfortunately it requires a dedicated fluorescence microscopy setup that is not available to all researchers in the field. Hence we wanted to develop an *in-situ* method that is based on standard equipment and easy to implement. Moreover we decided to refrain from measuring the length distribution, but rather aimed at developing a method for measuring the average length, which is sufficient for many practical purposes. This led us to a rather simple method based on electric-field induced differential absorption spectroscopy (EFIDAS), which allows determining directly the average length of SWCNTs in dispersions and provides chiral-index-resolution as an added value. The method takes advantage of the electric-field induced length-dependent dipole moment of nanotubes. In fact, electric-field induced alignment of SWCNTs in dispersion is not new and has been reported before.[14–16] However previous works in aqueous media could only resolve alignment of bundles of SWCNTs and not of well dispersed individual SWCNTs. Using a low-k solvent instead of water, it is in fact possible to resolve the field alignment of individually dispersed SWCNTs as we will show in this work. Moreover, since we have used dispersions with only a small number of different SWCNT chiral indices, it was possible to resolve the chiral index dependent average length. Besides opening a new approach for fast direct length mapping of dispersed SWCNTs, our method also gives novel insights into the concept of surfactant-induced nanotube conductance which was introduced some time ago to explain dielectrophoresis of semiconducting SWCNTs.[17] Furthermore simulations will show that the determination of the chiral-index resolved length distribution seems to be possible.

## RESULTS AND DISCUSSION

First we introduce the concept of differential absorption and its dependence on the electric field strength, on the SWCNT diameter and on the SWCNT length, before going on to discuss the experimental results. We start with the derivation of the electric-field induced differential absorption by considering the optical absorption of carbon nanotubes in between two parallel linear polarizers $A_\parallel$, modelled as[18,19]

$$A_\parallel = \sum_i \zeta_i \times N_i \times \frac{2 S_{3D,i} + 1}{3} \quad (1)$$

where $i$ denotes the different (n, m) SWCNT species in solution with absorption coefficient $\zeta_i$ and counting number $N_i$. $S_{3D}$ stands for the three-dimensional nematic order parameter,[19] which has a lower-limit of $S_{3D} = 0$ for a completely disordered phase, $S_{3D} \neq 0$ for a weakly ordered phase and an upper-limit of $S_{3D}=1$ corresponding to a perfectly ordered phase. For SWCNTs with diameter $d_{CNT}$ and length $l_{CNT}$ exposed to a homogeneous electric field $\vec{E}$ with angular frequency $\omega$, the nematic order parameter can be expressed as $S_{3D}=S_{3D}(|\vec{E}|,\omega,l_{CNT},d_{CNT},\varepsilon^*_{CNT},\varepsilon^*_\ell)$, with the complex dielectric functions $\varepsilon^*_{CNT}$ and $\varepsilon^*_\ell$ of the nanotubes and the solvent medium, respectively[20] (see also Supporting Information).

$S_{3D}$ is a direct measure of the orientation of SWCNTs under an external electric field, provided that the absorption coefficient $\zeta_i$ in equation (1) is constant under the applied electric field. For our experiment this assumption is valid when considering the electro-absorption experiments of surface-pinned SWCNTs by Izard et al.[21,22] Izard's data shows that upon applying an electric field of 12.5 kV/cm the changes of $\zeta_i$ are on the order of order of $10^{-4}$, whereas in our experiments with dispersed SWCNTs we observe changes in $A_\parallel$ on the order of $10^{-1}$ at an electric field of 2.5 kV/cm, as described later. Thus in our experiments $A_\parallel$ is predominantly determined by the orientation of SWCNTs as reflected by $S_{3D}$. Also transverse polarized transitions can be ignored due to a large depolarization effect in SWCNTs[23] The electric-field induced, differential absorption $\Delta A$ is then simply the relative difference of the optical absorption with and without an electric field. Since $\zeta_i$ is not electric field dependent, $\Delta A$ is only proportional to $S_{3D}$,

$$\Delta A = \frac{A_\parallel(E) - A_\parallel(E=0)}{A_\parallel(E=0)} = \frac{\left(\frac{2S_{3D}(E)+1}{3}\right) - \frac{1}{3}}{1/3} = 2S_{3D}(E) \qquad (2).$$

$S_{3D}$ can be expressed in terms of the alignment angle $\theta$ as,

$$S_{3D} = \frac{1}{2}(\overline{3\cos^2\theta_i} - 1) = \int \frac{1}{2}(3\cos^2\theta - 1)\cdot f(\theta, U_{ROT}, T)d\Omega \qquad (3).$$

The bar indicates the mean value, $\theta$ is measured between the long-axis of the nanotube and the electric field direction, which is also the polarization direction of the incident light in our experiment, and $\Omega$ is the solid angle. $U_{ROT}$ is the rotational energy of the SWCNT in the electric field and $f(\theta, U_{ROT}, T)$ is the Boltzmann distribution function. $U_{ROT}$ is calculated on the basis of the polarizability of SWCNTs in solution as described in detail by Blatt et al[24]. $T$ is the temperature and set to 300K. On this basis $S_{3D}$ and $\Delta A$ can be calculated as a function of $|\vec{E}|$, $\omega$, $l_{CNT}$, $d_{CNT}$, $\varepsilon^*_{CNT}$, $\varepsilon^*_l$ as outlined in the supporting information. $\varepsilon^*$ can be written as $\varepsilon^* = \varepsilon + i\sigma/\omega$, with $\varepsilon_\ell = 2.38\varepsilon_0$ and $\sigma_\ell = 1.0 \times 10^{-11} S/m$ for toluene.[25] For SWCNTs, $\varepsilon_{CNT}(d_{CNT})/\varepsilon_0 = 33.8 + 3.28 nm^2/d^2$ deduced from first-principles calculations of Marzari et al.,[26] yielding values between 35.4 $\varepsilon_0$ and 36.3 $\varepsilon_0$ for the SWCNTs under consideration. From previous dielectrophoresis experiments $\sigma_{CNT}$ is on the order of $0.1 S/m$.[17,24] It should be emphasized that the simulation contains no additional free parameter.

Figure 1 shows how the differential absorption of semiconducting SWCNTs in toluene depends on the nanotube length and diameter when exposed to an electric field $E$ = 2.5 kV/cm at a frequency $\nu = \omega/2\pi$ = 20 kHz. For SWCNTs with length up to 1.5 µm and diameter between 1-1.5 nm, $\Delta A$ can reach values of unity and therefore is experimentally accessible. Perfect alignment would correspond to $\Delta A = 2$, however this value cannot be reached due to depolarization effects, which become more pronounced for smaller diameter SWCNTs and will be discussed later. The sensitivity of $\Delta A$ to a certain SWCNT length range can be tuned by varying $E$ (Figure S4).

We will now compare the simulations with the experimental results. For measuring $\Delta A$ we have used the set-up shown in Figure 2. The core is a cuvette with external electrodes driven by a signal generator to generate an electric field that is perpendicular to the light path and parallel to the horizontal alignment of the two linear polarizers. Light from a halogen lamp is guided through the polarizers and cuvette, and detected by a silicon CCD based spectrometer. Each measurement of $\Delta A$ comprises a zero-field measurement and a field measurement. The signal generator supplies a 20 kHz signal with a voltage amplitude of 2380 V, resulting in an electric field strength in toluene of 2.5 kV/cm. This field strength has been used for the simulation in Figure 1. We have studied dispersions of polymer-wrapped SWCNTs in toluene containing mainly (9, 8), (10, 8), (10, 9), (11, 7), (11, 9), (11, 10) and (12, 8) SWCNTs detectable by the EFIDAS setup *via* their S22 transitions, as confirmed by additional absorption and photoluminescence spectroscopy (Figure S2, S3). We have measured EFIDAS with 4 fractions that were length sorted as described in [27]. For reference their length distribution was characterized by atomic force microscopy (AFM). Fitting the measured length distribution to a Gaussian and lognormal distribution function, the Gaussian mean value and lognormal peak values of the fractions were determined within the given uncertainty for fraction 1 to 790±18 nm and 720±27 nm, for fraction 2 to 543±19 nm and 465±24 nm, for fraction 3 to 564±19 nm and 480±39 nm, and for fraction 4 to 390±18 nm and 345±20 nm, respectively. Hence the average length of SWCNTs decreases from fraction 1 to fraction 4, which is a consequence of the longer retention time of shorter nanotubes on the column.[28] More details about the spectroscopy and materials are contained in the methods section.

The EFIDAS setup allows measuring $\Delta A$ associated with the second optical transitions S22 of the dispersions. Figure 3 shows the differential absorption spectra $\Delta A$ derived from the absorption spectra with and without applied electric field for the fractions 1-4 in (a, c, e, g), next to the corresponding length distributions determined by AFM in (b, d, f, h). First, we observe that the overall minimum and maximum of $\Delta A$ is 0.1 and 0.75, respectively, and hence within the range of the simulation. Second, we observe that $\Delta A$ scales systematically with the SWCNTs average length, with the highest value measured for fraction 1 and the lowest for fraction 4. This observation shows that EFIDAS is sensitive to the length of SWCNTs in dispersion and shows qualitatively a dependence that is expected based on the simulation. Third, we observe that $\Delta A$ is wavelength dependent, with maxima and minima correlating to structures in the absorption spectra. Figure 4a shows for fraction 1 an assignment of the peaks in $\Delta A$ to the SWCNT chiral index based on reference absorption and photoluminescence spectra shown in the supporting information. Since concentration effects drop out in EFIDAS we can trace back the modulations in $\Delta A$ to a chiral-index dependent degree of alignment. Interestingly the modulations in $\Delta A$ are also present in the shorter SWCNT fractions 2-4, where, according to Figure 1, $\Delta A$ is no longer sensitively dependent on the SWCNT chiral index or diameter. Hence a still observable modulation of $\Delta A$ must reflect a heterogeneous length distribution across the chiral index specific SWCNT subpopulations in the different fractions.

We will next extract the length of SWCNTs $l_{CNT}$ in dispersion from the differential absorption spectra $\Delta A$. In Figure 4a we plot $\Delta A$ for the fractions 1-4. Ranges of $\Delta A$ are indicated for each fraction by vertical bars. The experimental $\Delta A$ values can be converted into $l_{CNT}$ by using the simulation-derived correlation between $\Delta A$ and $l_{CNT}$ as shown in Figure 4b. The values of $l_{CNT}$ measured by EFIDAS are close to the AFM mean length and reproduce the trend in which higher fraction numbers have on average shorter SWCNTs (Figure 4c). As shown, for all fractions the EFIDAS derived values fit very well to the AFM mean length.

The agreement between the two methods is remarkable in particular since EFIDAS is based only on the dielectric functions of the SWCNTs and the solvent, and hence is without free parameters. Critical to the methods quantitative applicability is certainly the correlation between $\Delta A$ and $l_{CNT}$. It is interesting to see that the calculated curves in Figure 4b do not fall on top of each other. This is due to small variations in $\varepsilon_{CNT}$ and $d_{CNT}$ for the chiralities of interest. The influence of $\sigma_{CNT}$ is shown in Figure S5. For the simulation of Figure 4b and Figure 1 we have used $\sigma_{CNT} = 0.12 S/m$, a value that gives the best fit to the AFM data and which happens to be comparable to our earlier dielectrophoresis experiments with SWCNTs in aqueous medium.[24] Previously the non-zero value of $\sigma_{CNT}$ has been attributed to surface-induced conductivity mediated by ions in the aqueous double layer. For polymer-wrapped undoped SWCNTs in toluene as is the case here, we would have expected that $\sigma_{CNT}$ would be negligibly small. However simulations in Figure S5 show that for $\sigma_{CNT} = 0.1 S/m$ the maximum value for $\Delta A$ does not exceed 0.6, which is in contradiction with the experimental observations of (10, 8) SWCNTs, yielding a peak value of 0.63 from fraction 1. On the other hand a significantly larger value would shift the curve towards shorter $l_{CNT}$ and lead to an underestimation of the true length when benchmarked with the AFM measurements. Since surface-induced conductivity mediated by ions can be excluded in this work, $\sigma_{CNT}$ must be related to intrinsic free charge carriers present in the SWCNT. We attempt to estimate $\sigma_{CNT}$ by considering 4 conduction channels that are thermally activated by $\exp(-\Delta/2kT)$. With a transport gap of $\Delta$ = 1 eV[29] and $kT$ = 0.025eV, this would correspond to a thermally activated resistance of $R \approx 3 \cdot 10^{12}$ Ω. Assuming $l_{CNT} \approx 1 \mu m$ and $d_{CNT} \approx 1 nm$, $R$ can then formally be converted into $\sigma = 4l_{CNT}/\pi R d_{CNT}^2 \approx 0.4$ S/m, yielding the right order of magnitude. This oversimplified approach should be further developed including the influence of doping, however this would be beyond the scope of this paper.

We continue with the length of chiral index subpopulations encoded in the EFIDAS data, and extract from Figure 4a the differential absorption for different (n,m) species. The $\Delta A(n,m)$ values are converted into $l_{CNT}(n,m)$ and plotted against the fraction number as shown in Figure S6. The graph shows that overall the SWCNT length decreases with fraction number. Interestingly, we observe that according to EFIDAS the (n,m) length order changes from fraction 1 to fraction 2, which would mean that the elution time in GPC-based sorting not only depends on the SWCNT length but also on the chiral index. Whether the EFIDAS results are significant or whether the measurement error is underestimated remains currently unresolved. Unfortunately we cannot confirm this result by AFM measurements since AFM is not sensitive to the chiral index, but the data indicates the potential of EFIDAS for chirality-resolved length measurements in solution.

Now we discuss the dependence of EFIDAS on the electric field frequency, which is essential to understand the non-monotonic length dependence of $\Delta A$ for small diameter nanotubes in Figure 1. Effects related to the viscosity of the liquid can be ignored since we measure $\Delta A$ under steady-state conditions. The steady state is reached within seconds after switching on the electric field, as shown in Figure S7. Also nanotube-nanotube interaction can be ignored since the nanotube concentration is ~ 0.10 tube/µm³ as determined from the absorption in combination with the average nanotube length. As explained before, $\Delta A$ is calculated from the rotational energy $U_{ROT}$ using equations 2 and 3. $U_{ROT}$ is defined as the integral over the time-averaged torque $\langle T \rangle$ imposed on the SWCNTs in the oscillating field (eq. 4), whereas $\langle T \rangle$ is proportional to the difference of the real parts of the longitudinal and transverse Clausius-Mossotti factors $CM_{\parallel}$ and $CM_{\perp}$, respectively (eq. 5). The Clausius-Mossotti factors contain the frequency-dependent dielectric functions of the CNT and the solvent (eq. 6):

$$U_{ROT} = \int \langle T \rangle_{CNT} d\theta \tag{4}$$

$$\langle T \rangle_{CNT} = \vec{E}_{RMS}^2 \frac{\pi}{6} l_{CNT} d_{CNT}^2 \varepsilon_l \left[ \text{Re}\{CM_{\parallel}\} - \text{Re}\{CM_{\perp}\} \right] \sin\theta \cos\theta \tag{5}$$

$$CM(\omega)_{\parallel,\perp} = \frac{\varepsilon_{\parallel,\perp}^{CNT*}(\omega) - \varepsilon_{\ell}^{*}(\omega)}{\varepsilon_{\ell}^{*}(\omega) + (\varepsilon_{\parallel,\perp}^{CNT*}(\omega) - \varepsilon_{\ell}^{*}(\omega))L_{\parallel,\perp}} \tag{6}$$

$L_{\parallel}$ and $L_{\perp}$ are the longitudinal and transverse depolarization factors, respectively, with $L_{\perp}$ = 1/2 and $L_{\parallel} = d_{CNT}^2/l_{CNT}^2 \left[ \ln(2l_{CNT}/d_{CNT}) - 1 \right]$ .[17,30] The expressions can be simplified for semiconducting SWCNTs since $\varepsilon_{\parallel}^{CNT} = \varepsilon_{\perp}^{CNT} = \varepsilon_{CNT}$ and $\sigma_{\parallel}^{CNT} = \sigma_{\perp}^{CNT} = \sigma_{CNT}$, as outlined by Blatt et al.[24] Figure S8a shows that $\text{Re}\{CM_{\parallel}\} \gg \text{Re}\{CM_{\perp}\}$ throughout our simulation space. Hence the frequency dependence of $U_{ROT}$ and thus of $\Delta A$ is entirely determined by $\text{Re}\{CM_{\parallel}\}$. In Figure 5a we have plotted $\text{Re}\{CM_{\parallel}\}$ as a function of field frequency and SWCNT length for $d_{CNT}$ = 1.30 nm. The data shows that $\text{Re}\{CM_\parallel\} > 10^3$ for $\nu < 10^5$ Hz. Therefore, $\Delta A$ reaches sizable values only for $\nu < 10^5$ Hz as shown in Figure 5b. The experimental frequency $\nu = 2\cdot10^4$ Hz is hence a good choice for EFIDAS based length measurements of semiconducting SWCNTs in toluene, although larger $\Delta A$ values are expected for lower frequencies as shown in Figure 5b. A practical measure for selecting an appropriate field frequency is the so-called Maxwell-Wagner relaxation time $\tau_{MW}$, which accounts for the characteristic time scale of charge accumulation at the interface of two materials. For SWCNTs modelled as prolate ellipsoids,[24,30,31] the relevant longitudinal Maxwell-Wagner relaxation $\tau_{MW}^\parallel$ is given by $\tau_{MW}^\parallel = L_\parallel \varepsilon_{CNT} + (1-L_\parallel)\varepsilon_l / L_\parallel \sigma_{CNT} + (1-L_\parallel)\sigma_l$.

The corresponding Maxwell-Wagner relaxation frequency $(\tau_{MW}^\parallel)^{-1}$ decreases with the SWCNT length and increases with the SWCNT diameter, as shown in Figure S8b, and can be traced back to the diameter and length dependence of $L_\parallel$. A necessary condition for effective EFIDAS measurements is fulfilled if $\nu < (\tau_{MW}^\parallel)^{-1}$. Also the non-monotonic length dependence of $\Delta A$ for small diameter semiconducting SWCNTs in Figure 1 is due to the length and diameter dependence of $(\tau_{MW}^\parallel)^{-1}$ and Re{CM$_\parallel$}.

A limitation of EFIDAS is that the method does not work for semiconducting SWCNTs in aqueous surfactant solution which we discuss now and show experimental results. The conductivity of water with 1 wt-% SDS is $\sigma_{water}$ = 0.1 S/m, and 10 orders of magnitude larger compared to toluene. This large conductivity of the solution in combination with the dielectric constant of water ($\varepsilon_{water} = 81\varepsilon_0$) yields for the same applied voltage amplitude of $V_{rms}$ = 2380 V an electric field of only $E_{rms}$ = 1.2·10$^{-4}$ kV/cm at $\nu$ = 20 kHz, which is 5 orders of magnitude lower than in toluene. As a consequence of the very low field amplitude, $\Delta A$ is more than 10 orders of magnitude smaller as compared to toluene, and hence not detectable as calculated in figure 6a using $\sigma_\parallel^{CNT} = \sigma_\perp^{CNT} = 0.35 S/m$.[17,24] The corresponding experimental data is shown in Figure 6b for (8, 6) SWCNTs in 1 wt-% SDS in water. No significant values for $\Delta A$ could be measured. Since $E_{rms}$ in water increases with $\nu$ as shown in Figure S9, we simulated the dependence of $\Delta A$ on field frequency and SWCNT length for semiconducting SWCNTs in aqueous surfactant solution. The results are shown in Figure 6a and demonstrate that even at $\nu = 10^7$ Hz, where $E_{rms}$ in water reaches $5.3 \cdot 10^{-2}$ kV/cm, $\Delta A$ is on the order of $10^{-6}$ and therefore still too small to measure.

Finally we evaluate the potential of EFIDAS to determine not only the chiral-index resolved mean length of a dispersion but also the (n,m)-specific length distribution $g_{n,m}(l_{CNT})$. This requires measurement of $\Delta A_{n,m}$ as a function of the electric field $E$. $\Delta A_{n,m}$ is given by $\Delta A_{n,m}(E_{rms}) = \int g_{n,m}(l_{CNT}) \cdot f_{n,m}(l_{CNT}, E_{rms}) dl_{CNT}$, with $f_{n,m}(l_{CNT}, E_{rms}) = \Delta A_{n,m}(l_{CNT}, E_{rms})$ being the field $E_{rms}$ and length $l_{CNT}$ dependent (n,m)-specific differential absorption. The function $f_{n,m}$ is a result of the modelling as shown in Figure 7(a) and effectively maps the length distribution onto the field dependence. For demonstration purposes we use a log-normal model length distribution for $g_{n,m}(l_{CNT})$ as shown in Figure 7(b) and calculate $\Delta A_{n,m}(E_{rms})$ shown in Figure 7(c). The length distribution is encoded in $\Delta A_{n,m}(E_{rms})$ and can be extracted as follows. First we rewrite the above integral in vector notation $\Delta \vec{A} = \overline{F} \cdot \vec{g}$ with the vectors $\Delta \vec{A} = \Delta A_{n,m}(E_{rms})$ and $\vec{g} = g_{n,m}(l_{CNT})$, and the matrix $\overline{F} = f(l_{CNT}, E_{rms})$. We then numerically calculate the inverse matrix $\overline{F}^{-1}$, which finally allows us to reconstruct the length distribution via $\vec{g} = \overline{F}^{-1} \cdot \Delta \vec{A}$. The result is shown in Figure 7(b) and demonstrates that the model length distribution has indeed been nicely reconstructed by the outlined procedure. Notably, the accuracy of the approach is only limited by the size of $\overline{F}$ and the computational power required for calculating $\overline{F}^{-1}$. However, this is not a limiting factor and $\overline{F}^{-1}$ has to be calculated only once for each (n,m) SWCNT and can be further used as a look-up table. Hence we are convinced that EFIDAS allows to determine also the (n,m)-specific length distributions of SWCNTs in solution by measuring the electric-field dependent differential absorption spectrum. Testing EFIDAS with a set of monochiral dispersions should further corroborate these results.

# CONCLUSIONS

Electric-field induced differential absorption spectroscopy (EFIDS) is a fast, simple and cost-effective method to measure *in-situ* the average length of polymer-wrapped single-walled carbon nanotubes (SWCNTs) dispersed in toluene. Due to the spectral resolution of the method, it is possible to determine the average length of (n,m)-sub-populations in few-chiral index dispersions, which is particularly useful in the context of carbon nanotube processing as has been shown here for the case of length fractionation. The EFIDAS results are consistent with *ex-situ* atomic force microscopy (AFM) data obtained for the same fractions, which shows that the electric-field alignment of the nanotubes can be modelled on the basis of the dielectric properties of SWCNTs and toluene without free parameters. The method cannot be applied to SWCNTs dispersed in aqueous surfactant solution due to the high conductivity and permittivity of the solvent reducing the EFIDAS signal by orders of magnitude. However by simulations we could show that the determination of the chiral-index resolved *in-situ* length distribution of polymer-wrapped SWCNTs dispersed in toluene seems to be possible.

# MATERIALS AND METHODS

**SWCNT dispersions.** Dispersions of semiconducting SWCNTs (s-SWCNTs) in toluene were prepared from pulsed laser vaporization (PLV) SWCNTs[32] using the polymer Poly(9,9-di-n-dodecylfluorenyl-2,7-diyl) (PODOF). The PODOF wrapped s-SWCNTs in toluene were then further length fractionated with an GPC system as described in detail in [27]. The few-chiral index dispersions contain (10, 8), (10, 9), (11, 7), (11, 9), (11, 10) and (12, 8) SWCNTs. For comparison a dispersion enriched in (8, 6) SWCNTs in water was prepared using high pressure carbon monoxide (HiPco) SWCNTs from NanoIntegris and the surfactant sodium dodecyl sulfate (SDS) by size exclusion chromatography (SEC) as described in detail in [33].

**Spectroscopy.** EFIDAS measurements were performed in a home-made setup comprising a fiber-coupled Ocean Optics HR4000 High-Resolution Spectrometer, a fiber-coupled Mikropack DH-2000-BAL UV-VIS-NIR light source, Ocean Optics collimating lenses, Thorlabs LPVIS050 linear polarizers and a Hellma 114-QS Quartz cuvette with 12.5mm x 12.5mm outer dimensions and 4mm x 10mm inner dimensions. The cuvette was loaded with its long axis parallel to the light beam, yielding an optical path length of 10mm. Copper electrodes were attached from outside to the cuvette, parallel to the beam axis. The electrodes were biased with a Tesla generator dismantled from a commercial, low-value plasma ball. The generator has been characterized with a Testec TT HVP 15HF high-voltage probe and produces a 20 kHz signal with $V_{rms}$ = 2380 V when connected to the electrodes of a toluene filled cuvette. This voltage translates into an electric field of $E_{rms}$ = 2.5 kV/cm in toluene using the expression for a capacitive voltage divider

$$E_{rms}^{toluene} = \frac{1}{t_{toluene}} \cdot \frac{\varepsilon_{glass}/t_{glass}}{\varepsilon_{glass}/t_{glass} + 2\varepsilon_{toluene}/t_{toluene}} \cdot V_{rms},$$

with the thickness of the two quartz glass walls $t_{glass}$ = 4.25 mm, the thickness of the toluene layer $t_{toluene}$ = 4 mm, the dielectric constant of quartz glass $\varepsilon_{glass}$ = 3.75 and the dielectric constant of toluene $\varepsilon_{toluene}$ = 2.38. For simulations of the electric field in water ($\varepsilon_{water}$ = 81) the ionic conductance has been taken into account ($\sigma$ = 0.1 S/m for 1 wt-% SDS) yielding a frequency dependence of the field amplitude as shown in Figure S9. In the EFIDAS setup both polarizers were aligned parallel to the electric field axis. The spectral range of the setup is 300 - 1000 nm. In addition a Varian Cary 500 spectrometer was used for absorption measurements over an extended spectral range of 250 - 2000 nm (Figure S1), without electric fields. Photoluminescence excitation maps (Figure S2) were measured with a modified Bruker IFS66 FTIR spectrometer equipped with a liquid-nitrogen cooled Germanium photodiode and a monochromatized halogen lamp. The system has an excitation range of 700 - 950 nm and a detection range of 1250 - 1700 nm.[34]

**Atomic force microscopy.** The diameter and length distributions of PODOF-wrapped SWCNTs were measured with a Bruker NanoScope III atomic force microscope using Mikromash silicon cantilevers in tapping mode at a resonance frequency of 320 kHz. The samples were prepared by spin-coated 0.5 μl of dispersion at 4000 rpm for 1 min onto a silicon wafer.

**Calculations.** Calculations were performed using Matlab R2014a software.

*Conflict of Interest:* The authors declare no competing financial interest.

*Acknowledgment.* The authors acknowledge support from the Helmholtz Research Program Science and Technology of Nanosystems (STN). B. S. Flavel acknowledges support from the Deutsche Forschungsgemeinschafts Emmy Noether Program under grant number FL 834/1-1.

*Supporting information Available:* Additional information on the EFIDAS model (Fig. S1), Absorption spectra and fluorescence excitation maps of SWCNT solutions (Fig. S2,S3), Differential absorption vs. electric field and SWCNT length (Fig. S4), Differential absorption vs. SWCNT conductivity and length (Fig. S5), Chirality-resolved mean length vs. fraction number (Fig. S6), Simulation Clausius-Mossotti factor ratio Re $CM_{\parallel}$ / Re $CM_{\perp}$ and the longitudinal Maxwell-Wagner relaxation frequency (Fig. S8), Electric field *versus* field frequency in toluene and in water (Fig. S9). This material is free of charge *via* the Internet at http://pubs.acs.org.

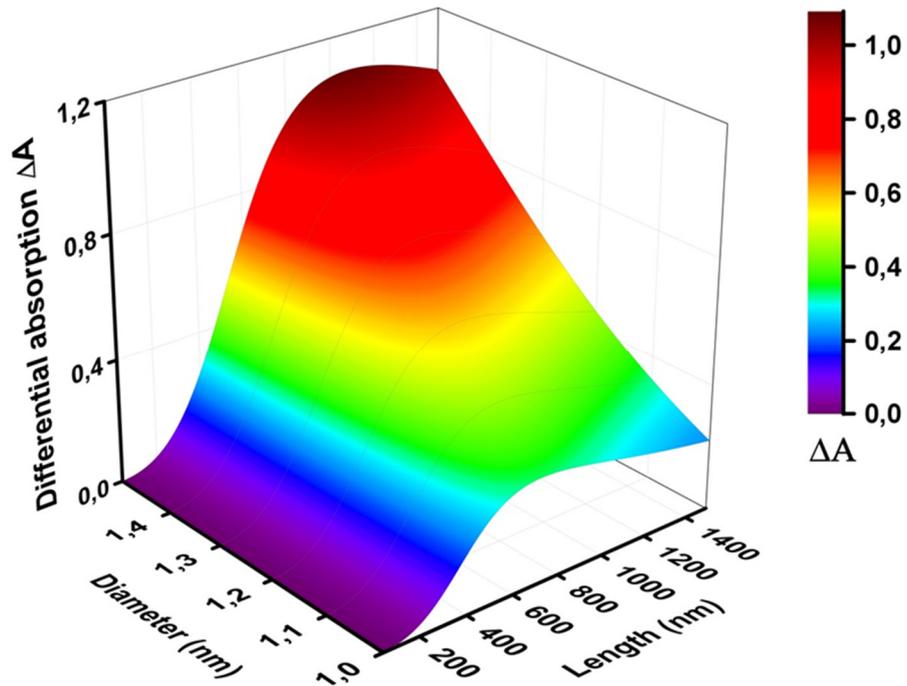

**Figure 1.** Simulation of differential absorption $\Delta A$ as a function of SWCNT diameter $d_{CNT}$ and length $l_{CNT}$ for semiconducting SWCNTs dispersed in toluene. The electric field and frequency was set to $E_{rms}$ = 2.5 kV/cm and $v$ = 20 kHz.

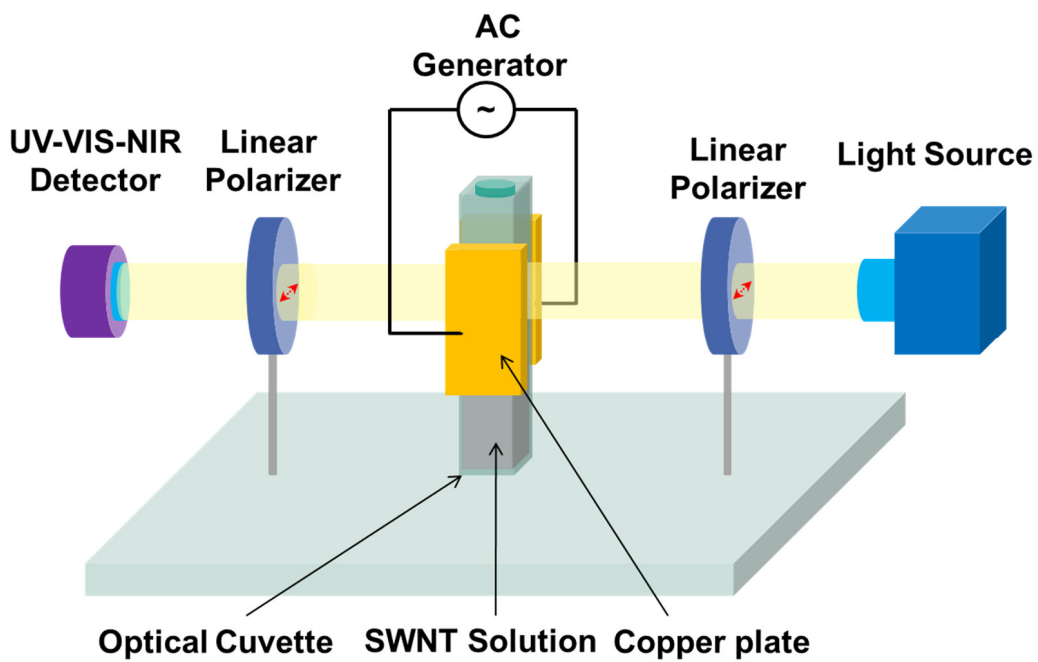

**Figure 2. Schematic of the electric-field induced differential absorption spectroscopy (EFIDAS) setup. The linear polarizers are horizontally aligned (red arrows), parallel to the direction of the electric field generated across the cuvette.**

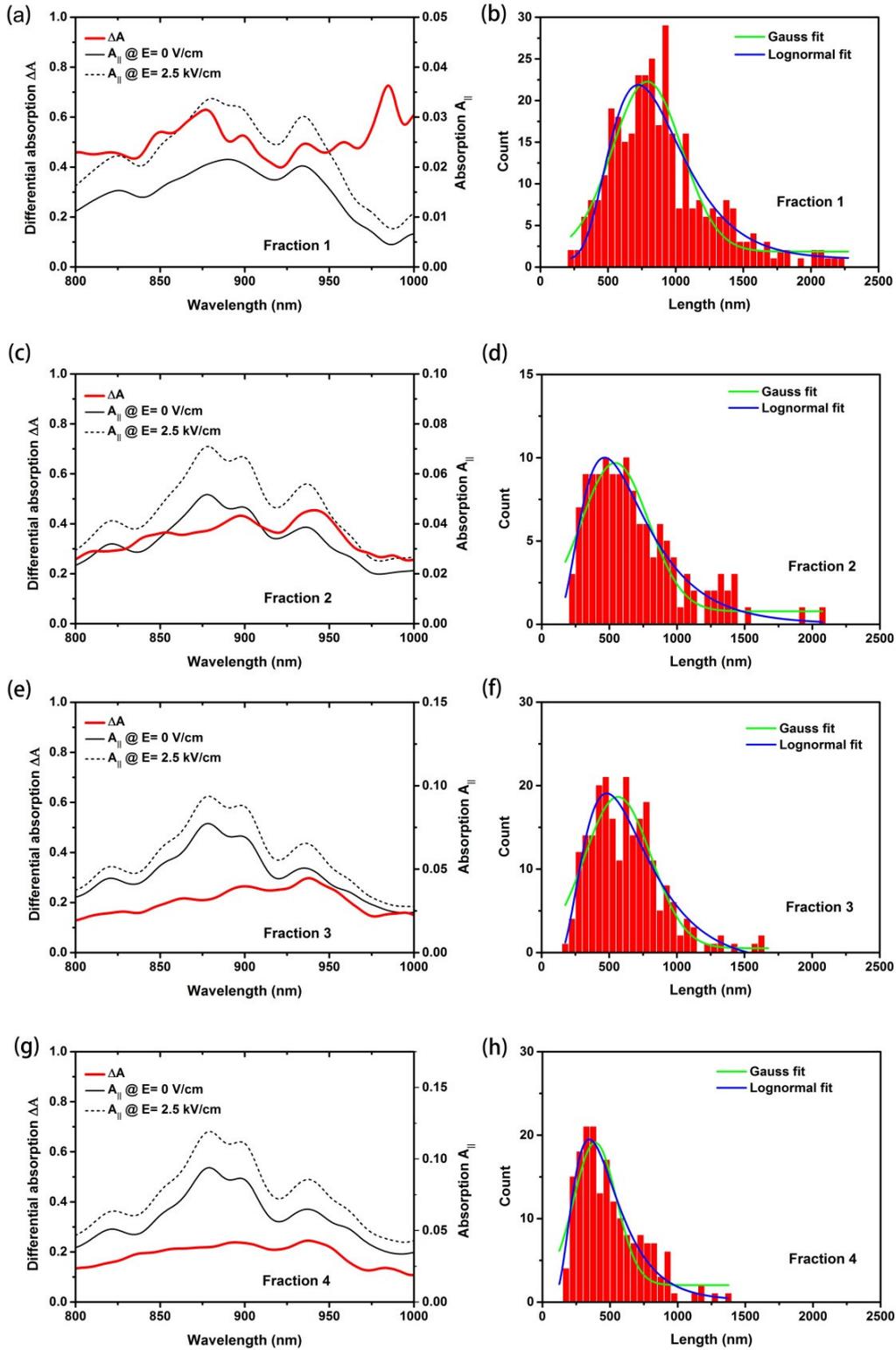

**Figure 3.** Differential absorption spectra $\Delta A$ and absorption spectra $A_{\parallel}$ of semiconducting SWCNTs in toluene, measured at zero field and at $E_{\text{rms}}$ = 2.5 kV/cm and $v$ = 20 kHz. The data is shown for fractions 1-4 (a, c, e, g) together with AFM length measurements (b, d, f, h).

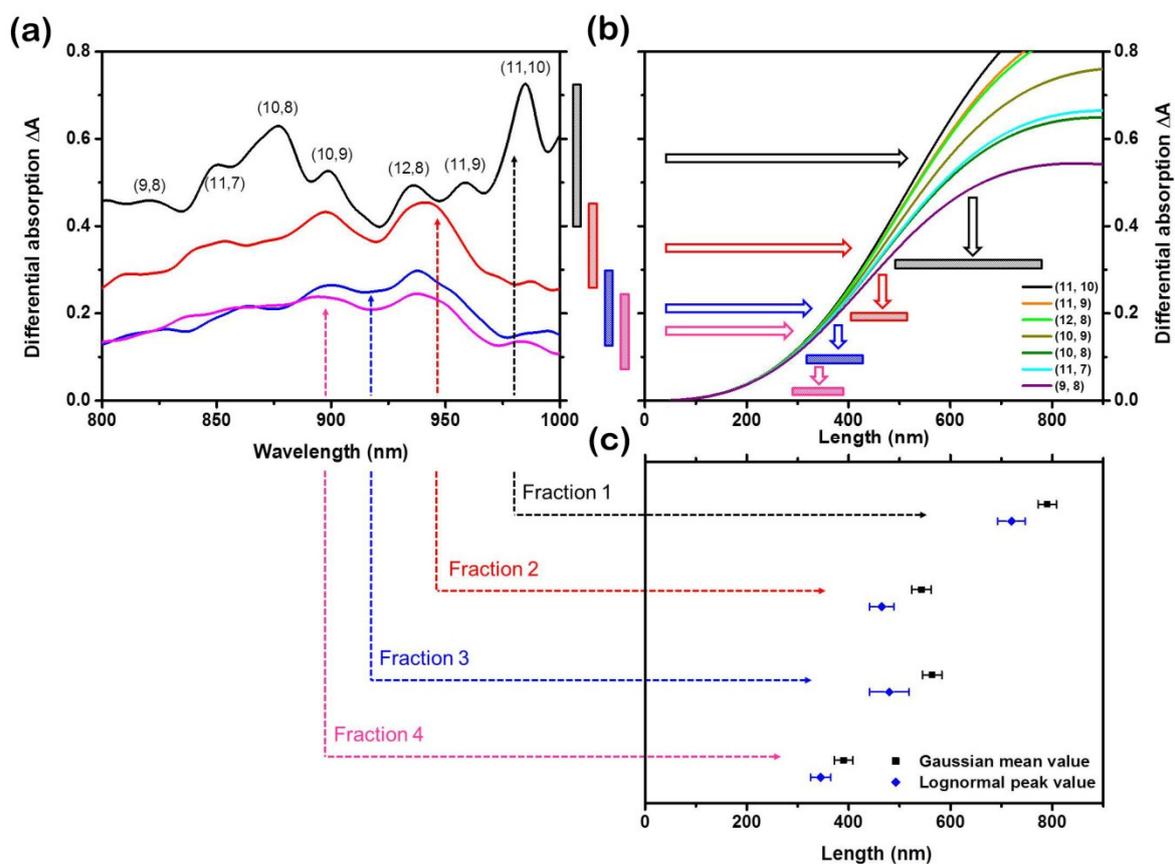

**Figure 4. Length determination with EFIDAS. (a) Differential absorption spectra $\Delta A$ measured for fractions 1-5 with indicated (n, m)-specific contributions. (b) (n, m)-specific calculations of $\Delta A$ *versus* nanotube length $l_{CNT}$. Ranges of $\Delta A$ are converted into ranges of $l_{CNT}$ as indicated by bars and arrows. The results are compared to fitted AFM mean and peak length values for the fractions shown in (c).**

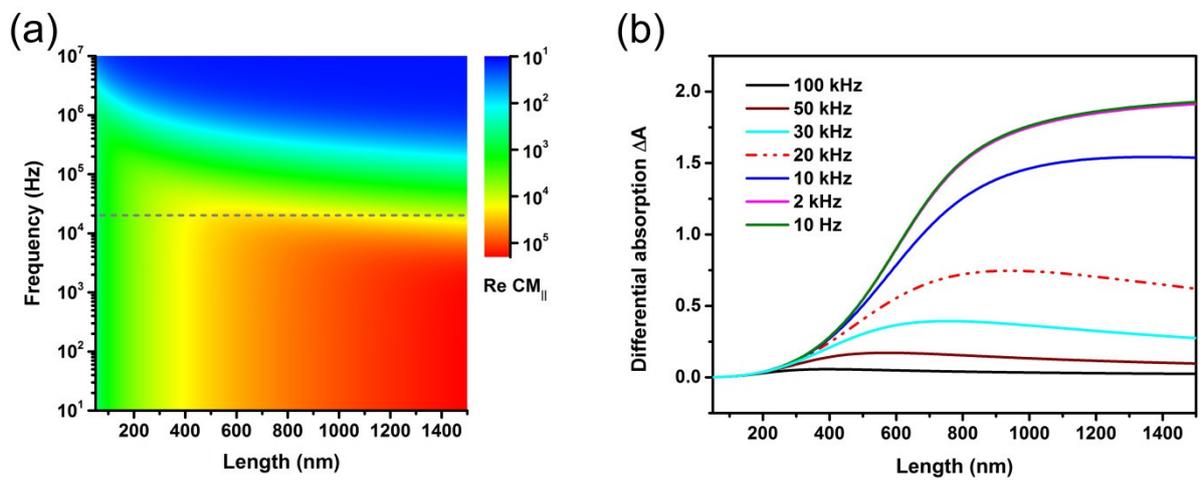

**Figure 5.** Simulations of (a) the real part of the longitudinal Clausius-Mossotti factor Re $CM_\parallel$ and (b) the differential absorption $\Delta A$, as a function of field frequency $\nu$ and nanotube length $l_{CNT}$. Simulations consider semiconducting SWCNTs with diameter $d_{CNT}$ = 1.30 nm, dispersed in toluene and exposed to an electric field $E_{rms}$ = 2.5 kV/cm. The horizontal dashed line in (a) indicates the field frequency used for the EFIDAS measurements.

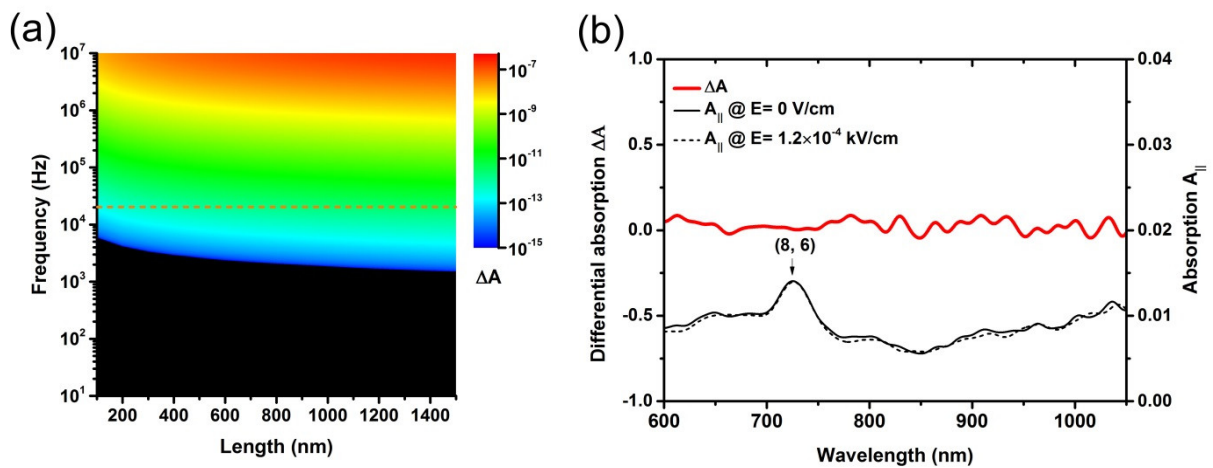

**Figure 6. EFIDAS with semiconducting SWCNTs in aqueous surfactant solution. (a)** Simulation of differential absorption $\Delta A$ as a function of field frequency $\nu$ and nanotube length $l_{CNT}$, for SWCNTs in 1-wt%-SDS in water. The horizontal dashed line indicates the field frequency used in the experiment. **(b)** Measurements of differential absorption spectra $\Delta A$ and absorption spectra $A_\parallel$ at zero field and at $E_{rms}$ = 1.2·10$^{-4}$ kV/cm and $\nu$ = 20 kHz. The electric field in water is reduced by 5 orders of magnitude and $\Delta A$ by 12 orders of magnitude as compared to toluene.

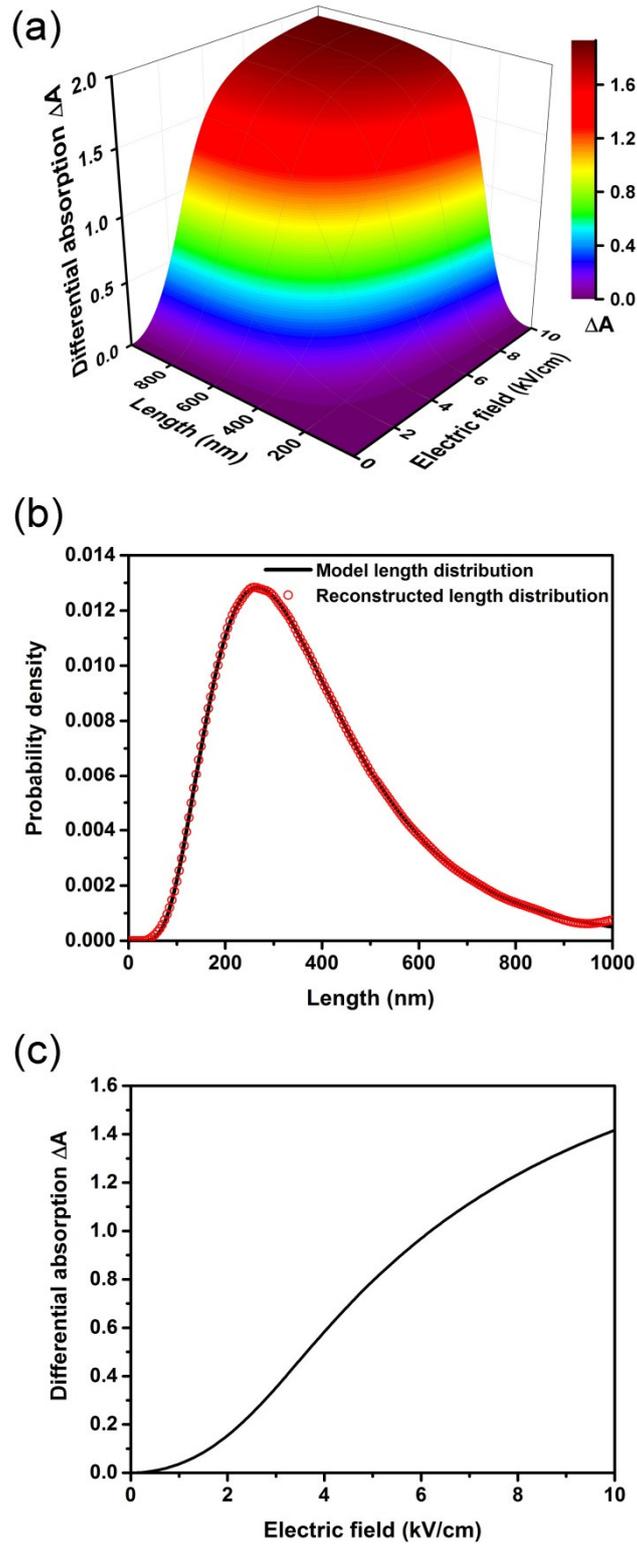

**Figure 7. Strategy to determine (n,m)-specific length distribution of SWCNTs in dispersion with EFIDAS. (a)** Simulation of the differential absorption ΔA as a function of SWCNT length $l_{CNT}$ and electric field $E_{rms}$, for $d_{CNT}$ = 1.30 nm and $v$ = 20 kHz. **(b)** Log-normal model length distribution (solid line) compared to the length distribution (open symbol) reconstructed from (c) as described in the text. **(c)** Differential absorption *versus* electric field $E_{rms}$ for the log-normal model length distribution shown in (b).

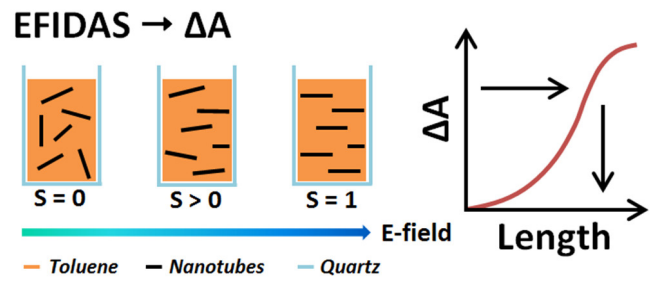

**Table of content figure**

# Chirality-Dependent Length Mapping of Carbon Nanotubes in Solution Using Electric-Field Induced Differential Absorption Spectroscopy

Wenshan Li[1,2], Frank Hennrich[1], Benjamin S. Flavel[1], Manfred M. Kappes[1,3], and Ralph Krupke[1,2*]

[1]Institute of Nanotechnology, Karlsruhe Institute of Technology, 76021 Karlsruhe, Germany
[2]Department of Materials and Earth Sciences, Technische Universität Darmstadt, 64287 Darmstadt, Germany
[3]Institute of Physical Chemistry, Karlsruhe Institute of Technology, 76128 Karlsruhe, Germany

email: wenshan.li@kit.edu; krupke@kit.edu

# Supporting Information

Additional information on the EFIDAS model (Fig. S1)

Absorption spectra and fluorescence excitation maps of SWCNT solutions (Fig. S2, S3)

Differential absorption vs. electric field and SWCNT length (Fig. S4)

Differential absorption vs. SWCNT conductivity and length (Fig. S5)

Chirality-resolved mean length vs. fraction number (Fig. S6)

Dynamic effects in EFIDAS (Fig. S7)

Simulation Clausius-Mossotti factor ratio Re $CM_{\parallel}$ / Re $CM_{\perp}$ and the longitudinal Maxwell-Wagner relaxation frequency (Fig. S8)

Electric field versus field frequency in toluene and in water (Fig. S9)

# Additional information on the EFIDAS model

In order to derive the differential absorption $\Delta A$, or the nematic order parameter $S_{3D}$ of SWCNTs in solution,[1] we need the Boltzmann distribution function distribution $f(\theta, U_{ROT}, T)$ which depends on the rotational energy $U_{ROT}$ and the temperature $T$ and is given by[2]

$$f(\theta)d\Omega = \frac{\exp(-U_{ROT}/k_B T)d\Omega}{\int_0^\pi \exp(-U_{ROT}/k_B T)d\Omega} \qquad (1),$$

$$d\Omega = \sin\theta\, d\theta\, d\varphi = 2\pi \sin\theta\, d\theta \qquad (2).$$

$f(\theta)d\Omega$ describes the probability for specific orientation of one SWCNT with respect to the electric field expecting in the unit radian angle of $d\theta$ in spherical coordinates as illustrated in Figure S1. $k_B$ is the Boltzmann constant.

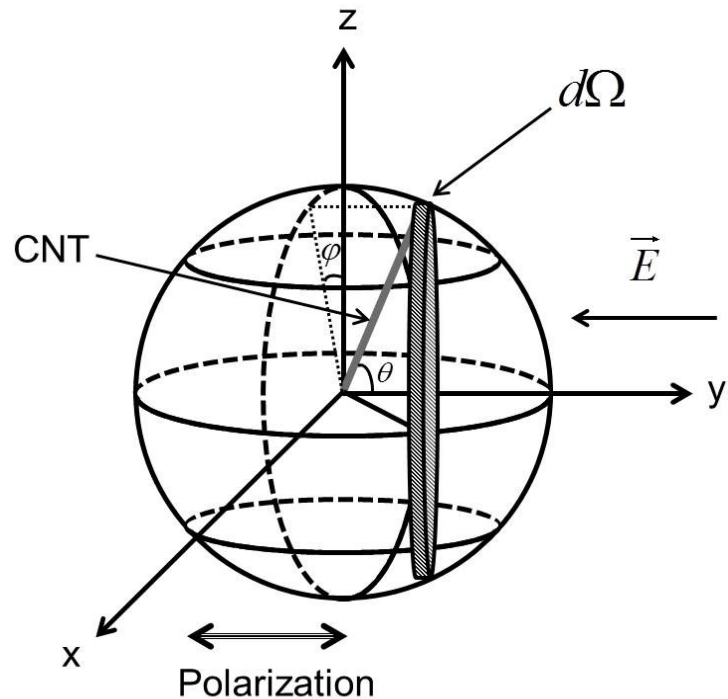

**Figure S1. Cartesian and spherical coordinates for the EFIDA spectroscopy-based model of SWCNTs, the direction of polarized light is parallel to the electric field.**

# Absorption spectra of SWCNT solutions

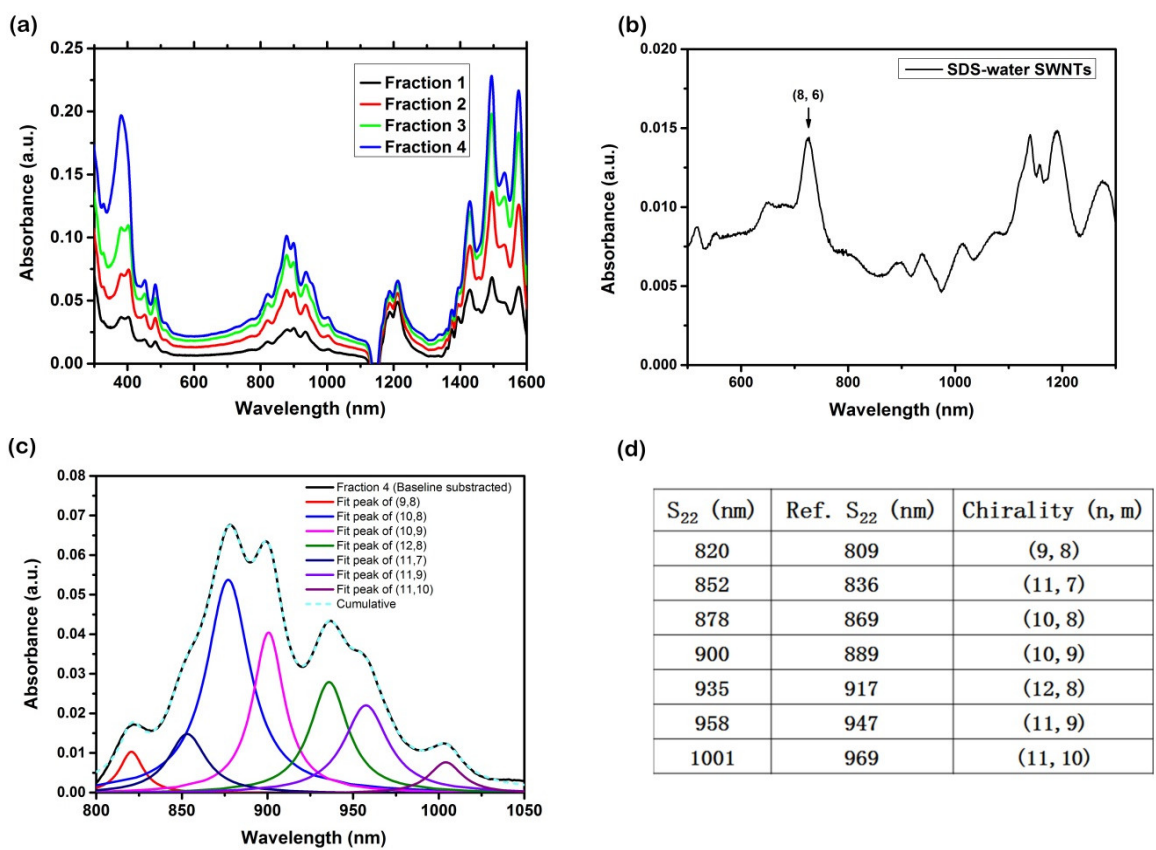

**Figure S2.** Absorption spectra of PODOF-wrapped semiconducting SWCNTs in toluene (a) and (8,6) dispersed in SDS/water (b). The data has been measured with a complementary setup that extends the range of the EFIDAS setup towards the NIR. (c) Zoom-in to the S22 region that is relevant for the EFIDAS measurements. The absorption data has been deconvoluted into individual (n,m) contribution using Lorentzian fit functions after subtracting a linear background. The dashed line is the cumulative of the individual contributions and fits very well to the data. (d) (n m)-assignment of the individual peaks based on reference data from SWCNTs in aqueous surfactant medium.[3] Differences in the peak positions are due to the different solvents (toluene versus water).

# Fluorescence excitation maps of SWCNT solutions

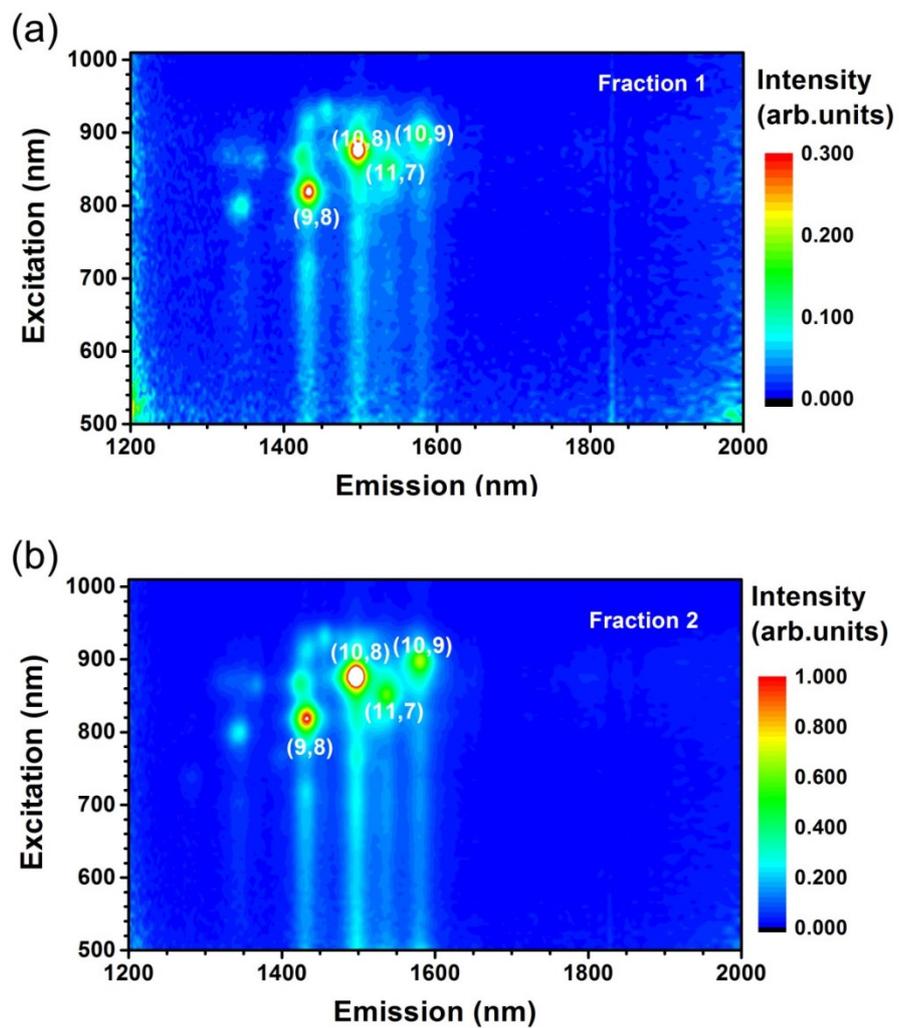

Figure S3. Photoluminescence excitation map of PODOF-wrapped semiconducting SWCNTs in toluene. (a) fraction 1, (b) fraction 2. CNT species with emission line above 1650 nm are not detectable due to the strong absorption of toluene between 1650-1700 nm.

## Differential absorption vs. electric field and SWCNT length

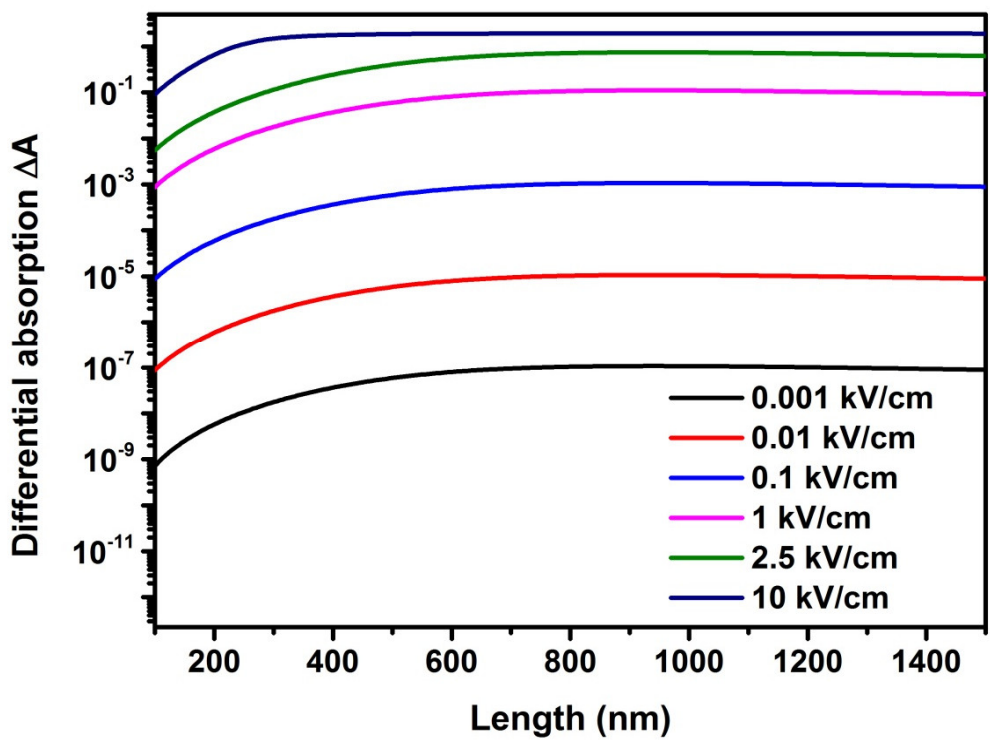

**Figure S4.** Differential absorption $\Delta A$ as a function of electric field strength field *E* and nanotube length $l_{CNT}$. Simulations were done for semiconducting SWCNTs with diameter $d_{CNT}$ = 1.30 nm, dispersed in toluene. The field frequency was set to $\nu$ = 20 kHz.

## Differential absorption vs. SWCNT conductivity and length

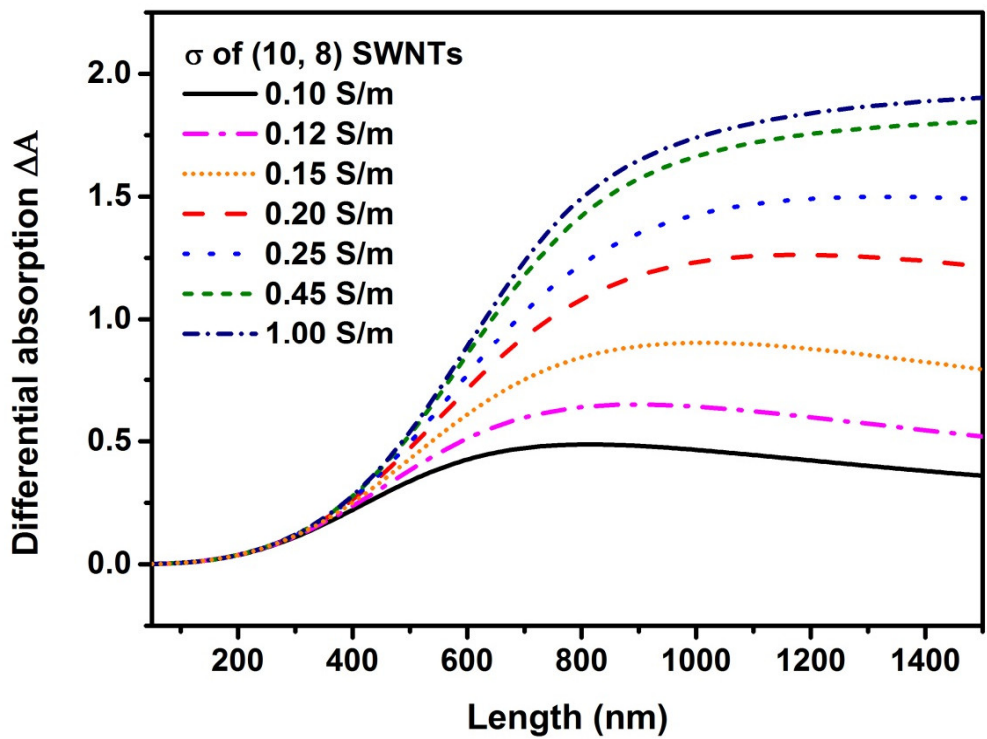

**Figure S5.** Differential absorption $\Delta A$ as a function of nanotube conductivity σ$_{CNT}$ and nanotube length $l_{CNT}$ for (10, 8) SWCNTs dispersed in toluene. The electric field and frequency was set to $E_{rms}$ = 2.5 kV/cm and v = 20 kHz.

## Chirality-resolved mean length vs. fraction number

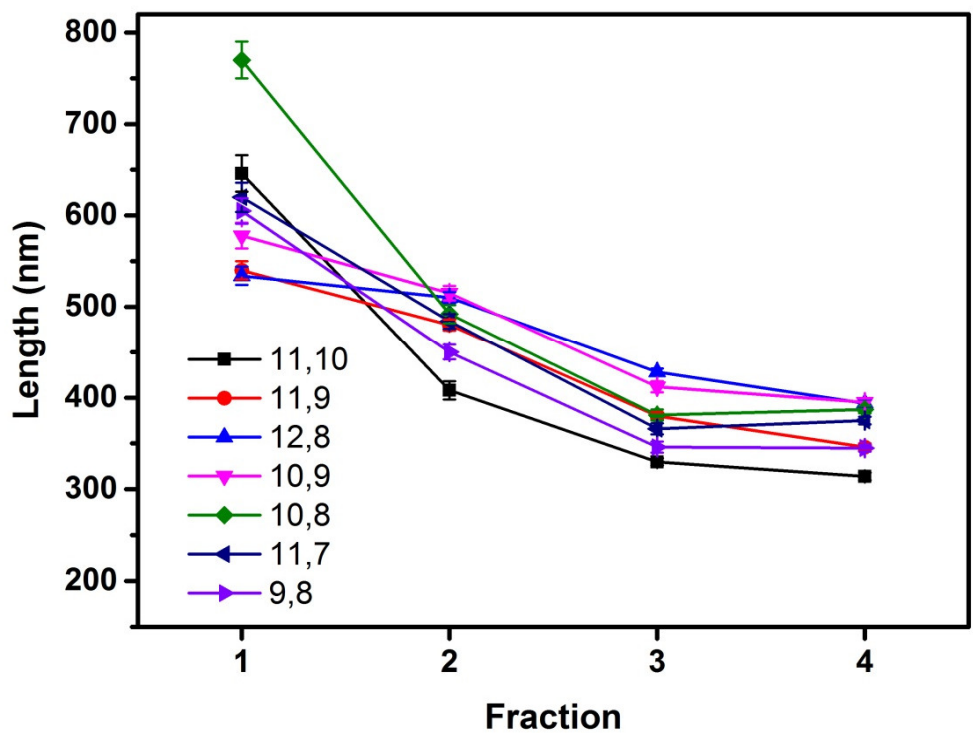

**Figure S6. Chirality-resolved mean length determined by EFIDAS for SEC-sorting fractions 1-4.**

## Dynamic effects in EFIDAS

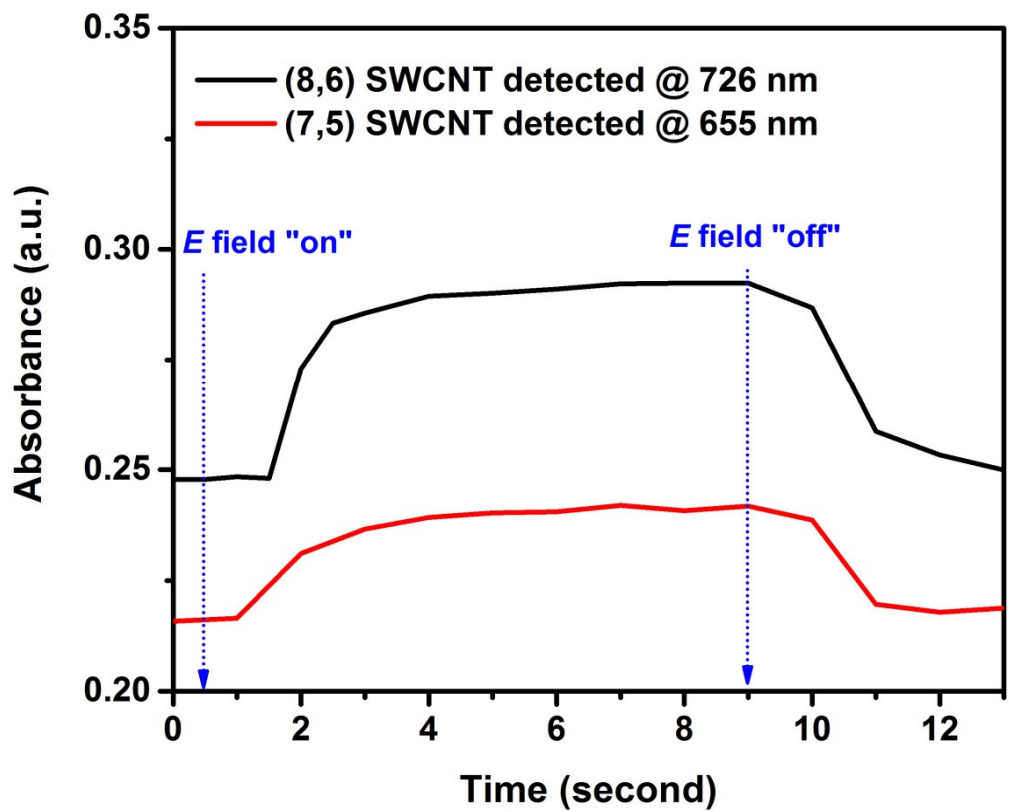

**Figure S7.** Dynamics of nanotube alignment measured by the time-evolution of the absorbance upon switching of the electric field..

# Simulation Clausius-Mossotti factor ratio Re CM$_{\parallel}$ / Re CM$_{\perp}$ and the longitudinal Maxwell-Wagner relaxation frequency

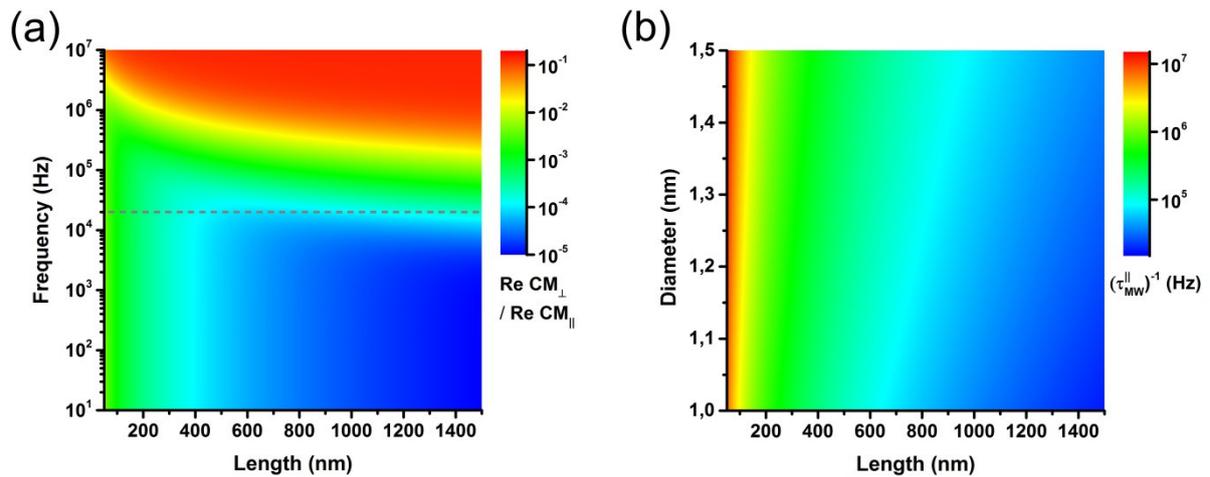

**Figure S8.** Simulation of (a) the ratio of the real part of the transverse and longitudinal Clausius-Mossotti factors Re CM$_{\parallel}$ / Re CM$_{\perp}$ as a function of field frequency $\nu$ and nanotube length $l_{CNT}$, and (b) the longitudinal Maxwell-Wagner relaxation frequency $\left(\tau_{MW}^{\parallel}\right)^{-1}$ versus nanotube length $l_{CNT}$ and nanotube diameter $d_{CNT}$. Simulations were done for semiconducting SWCNTs dispersed in toluene. In (a) $d_{CNT}$ = 1.30 nm and $E_{rms}$ = 2.5 kV/cm, and the horizontal dashed line indicates the field frequency used for the EFIDAS measurements.

## Electric field versus field frequency in toluene and in water

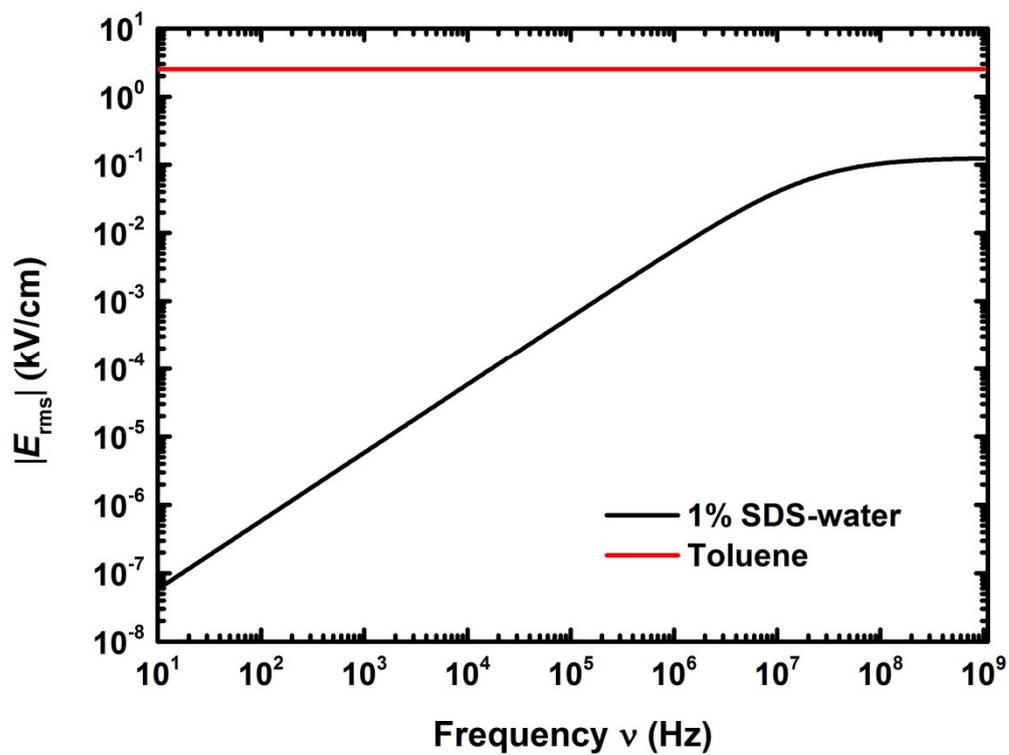

**Figure S9: Electric field $E_{rms}$ versus field frequency $\nu$ in toluene and in water with 1 wt-% SDS. The voltage amplitude applied to the cuvette is $V_{rms}$ = 2380 V in both cases.**